\documentclass[9pt,twocolumn,twoside]{opticajnl}
\journal{opticajournal} % use for journal or Optica Open submissions

\setboolean{shortarticle}{true}
\usepackage{slashbox}
\usepackage{mathtools}              %added by Emma
\usepackage{svg} %added by Emma
\usepackage{tabularx}
\usepackage{epstopdf}
\title{Certifying spatial entanglement between non-degenerate photon pairs with a camera}

\author[1,2,*]{Emma Brambila}
\author[3]{Raphael Guitter}
\author[1,4]{René Sondenheimer}
\author[1,5]{Markus Gräfe}
\author[3]{Hugo Defienne}

\affil[1]{Fraunhofer Institute for Applied Optics and Precision Engineering. Albert-Einstein-Straße 7. 07745 Jena, Germany.}
\affil[2]{Institute of Applied Physics, Abbe Center of Photonics, Friedrich Schiller University Jena. Max-Wien-Platz 1. 07743 Jena, Germany.}
\affil[3]{Sorbonne Université, CNRS, Institut des NanoSciences de Paris, INSP, F-75005 Paris, France.}
\affil[4]{Institute of Condensed Matter Theory and Optics, Friedrich Schiller University Jena. Max-Wien-Platz 1. 07743 Jena, Germany.}
\affil[5]{Institute for Applied Physics, Technical University of Darmstadt. Otto-Berndt-Str. 3, 64287 Darmstadt, Germany.}
\affil[*]{emma.celina.brambila-tamayo@iof.fraunhofer.de}

\begin{abstract}
We investigate transverse spatial entanglement between photon pairs of different wavelengths using a camera-based coincidence technique.
By adapting the correlation measurements to the photons frequencies, we certify the presence of entanglement between the pairs through violation of an Einstein–Podolsky–Rosen criterion. Additionally, we examine how parameters such as pump waist and crystal length influence these correlations. 
Our results highlight key differences from the frequency-degenerate case, showing that an adapted theoretical analysis is essential to avoid significant misestimations and to reliably certify entanglement. % 
\end{abstract}

\setboolean{displaycopyright}{false} % Do not include copyright or licensing information in submission.

\begin{document}

\maketitle

\section{Introduction}
Quantum correlations between photons that are non-degenerated in frequency are of great interest for many applications. 
In quantum imaging~\cite{Marta2019,defienne_advances_2024}, induced-coherence~\cite{wang_induced_1991,Lemos} and Ghost imaging~\cite{pittman_optical_1995} approaches take advantage of spatially correlated photon pairs at very different wavelengths. They allow for the study and illumination of samples at frequencies that are difficult to detect - such as in the mid-infrared and terahertz - while enabling imaging or spectrometric measurements using visible light detectors~\cite{aspden_photon-sparse_2015,GIBoyd,topfer_quantum_2022,fuenzalida_experimental_2023,vanselow_frequency-domain_2020,kviatkovsky_microscopy_2020,pearce_practical_2023,kalashnikov_infrared_2016,kutas_terahertz_2024,gilaberte_basset_experimental_2023}. 
% I do not understand the argument in this sentence, since in general both photons are degenerate: Furthermore, quantum communication protocols based on polarization entanglement can be designed to minimize losses by harnessing different wavelength ranges accordingly, for example, using telecom wavelengths for fiber transmission or near-infrared for free-space links~\cite{NRevHu}.
Furthermore, using frequency-separated and potentially broadband photons is also of interest for quantum communication~\cite{brecht_photon_2015,roslund_wavelength-multiplexed_2014,wengerowsky_entanglement-based_2018}.

\begin{figure}[h]
\centering
\includegraphics[width=1\linewidth]{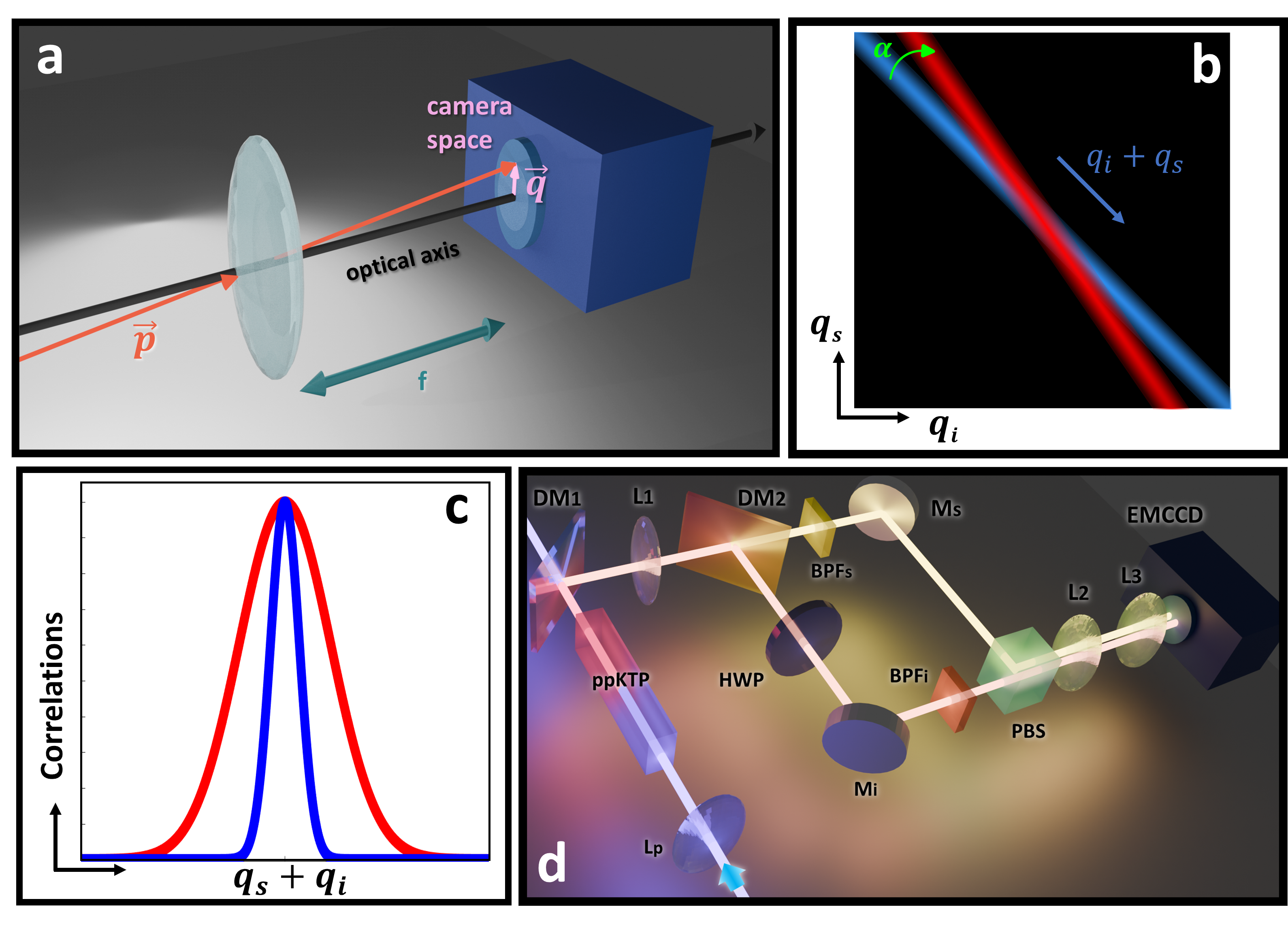}% Here is how to import EPS art
\caption{\label{fig:SetupFF} \textbf{a,} Illustration of the action of a lens with focal length $f$, which maps an optical wave with momentum $\vec{p}$ and wavelength $\lambda$ to its corresponding transverse position $\vec{q} = \frac{f \lambda}{2\pi\hbar} \vec{p}$ in the camera plane.
\textbf{b,} Illustration of the joint probability distribution (JPD) measured in the camera using a Fourier imaging configuration for degenerate (blue line) and non-degenerate (red line) photons (one dimension). \textbf{c,} Illustration of JPD projections along the sum-coordinate axis $q_s+q_i$ for degenerate (blue curve) and non-degenerate (red curve) photons. Due to the tilt by an angle $\alpha = \arctan(\lambda_s / \lambda_i)$, the momentum width measurement in the case of non-degenerate photons becomes not accurate because it appears broadened. To recover the true correlation width, the JPD must be projected along the corrected variable that accounts for this tilt, namely $q_s / \lambda_s + q_i / \lambda_i$. \textbf{d,} Camera-based coincidence detection setup for measuring momentum correlations between photons of different wavelengths. L denotes lenses, DM is a dichroic mirror, HWP a half-wave plate, M a mirror, and PBS a polarizing beam splitter. Bandpass filters (BPF) are placed in the signal and idler paths to spectrally narrow their respective wavelengths. The combination of lenses L$_1$–L$_3$ are used to image either the crystal surface or the Fourier plane of the crystal onto the detection plane. At the camera (EMCCD), two additional long-pass filters (not shown) are used to reduce background noise and to block any residual pump photons.}
\end{figure}

Spontaneous parametric down-conversion (SPDC) in nonlinear crystals enables the generation of spatially entangled photon pairs which can also be non-degenerate in frequency. To exploit this entanglement, it is often necessary to certify it across the many accessible spatial modes. While initially performed using scanning techniques~\cite{howell_realization_2004}, this process has been enhanced by the advent of single-photon-sensitive cameras. Electron-Multiplying Charge-Coupled Device (EMCCD) cameras~\cite{Padgett2012,moreau_realization_2012,lantz_einstein-podolsky-rosen_2015}, intensified Complementary Metal-Oxide-Semiconductor (iCMOS) cameras~\cite{dabrowski_einsteinpodolskyrosen_2017,dabrowski_certification_2018}, Single-Photon Avalanche Diode (SPAD) arrays~\cite{ndagano_imaging_2020,eckmann_characterization_2020}, and intensified time-stamping cameras such as the Tpx3Cam~\cite{courme_quantifying_2023,li_rapid_2025} have all been successfully used to measure and certify spatial entanglement.

However, all demonstrations of spatial entanglement characterization using cameras have so far been limited to wavelength-degenerate photons, most often at $810$ nm. In our study, we certify transverse spatial entanglement between non-degenerate photon pairs generated by SPDC using an EMCCD camera. By adapting the correlation measurements to account for the photons frequency difference, we accurately estimate their position and momentum correlations and demonstrate a violation of an Einstein-Podolsky-Rosen (EPR) criterion~\cite{reid_demonstration_1989,cavalcanti_experimental_2009,reid_colloquium_2009}. Crucially, we show that this violation would not be observed in using previous camera-based methods that do not take into account the frequency difference. We also investigate how the pump beam and crystal length affect these spatial correlations.

\section{Entanglement certification}

A practical way to certify transverse spatial entanglement with a camera is by violating an EPR criterion~\cite{reid_demonstration_1989,cavalcanti_experimental_2009} i.e. showing that
\begin{equation}
\label{eprreid}
\Delta_r \Delta_{p} \leq \frac{\hbar}{2},
\end{equation}
where $\hbar$ is the reduced Planck constant, $\Delta_r$ and $\Delta_p$ are the variances of the transverse position and momentum coordinates, respectively, along a single spatial axis ($x$ or $y$). Assuming a Gaussian structure of the two-photon wavefunction~\cite{fedorov_gaussian_2009,Baghi2022}, these quantities correspond to the widths of the joint transverse position and momentum probability distributions {(see supplementary document)}.
%\textcolor{red}{Emma: I think here you refer to eq. 39 from the reference, but there they explain that this is not valid for collinear emission (ppKTP case), so I think we need to take the full expression (eq. 40). This is what I wanted to explained in one part of the Appendix, because that can be reduced to a simpler gaussian when the wavelengths are close enough.}
 To certify entanglement, the spatial joint probability distribution (JPD) of photon pairs is first measured in both the position and momentum bases using appropriate lenses arrangements~\cite{Defienne}. The resulting distributions are then projected onto the minus-coordinate $(\vec{r}_s - \vec{r}_i)$ and sum-coordinate $(\vec{p}_s + \vec{p}_i)$ axes, where \(\vec{r}_{s/i}\) and \(\vec{p}_{s/i}\) denote the position and momentum of signal/idler photons, respectively. The values of \(\Delta_r\) and \(\Delta_p\) are extracted from the widths of the peaks observed of these projections using Gaussian fitting~\cite{fedorov_gaussian_2009}. For frequency-degenerate photon pairs, previous works using EMCCD cameras~\cite{moreau_realization_2012,Padgett2012} have demonstrated violations of the EPR criterion by several orders of magnitude.

In the approach described above, access to the momentum basis is achieved by arranging lenses to perform an optical Fourier transform. In such a configuration, an incoming photon with transverse momentum \(\vec{p}\) is mapped to a camera pixel of transverse position \(\vec{q}\) according to the formula
\begin{equation}\label{eq:Geo}
    \vec{q}= \frac{f \lambda}{2\pi\hbar} \vec{p},
\end{equation}
%\textcolor{red}{here we are talking about $\vec{k}=\vec{p}/\hbar$, not sure if that is a bit confusing, but now the experimental numbers ($\Delta k$) match with the units of $\Delta_{p}$ by adding the $\hbar$ everywhere.}
where $\lambda$ is the photon wavelength and $f$ the focal length (Fig.~\ref{fig:SetupFF}a). \eqref{eq:Geo} also shows that two incident photons with the same transverse momentum, \(\vec{p_s}=\vec{p_i}\), but different frequencies, \( \lambda_s \neq \lambda_i\), are not detected at the same position on the camera. For non-degenerate photon pairs, this wavelength-dependent scaling effect would introduce a systematic error in the measurement of \(\Delta_p\) if not accounted for properly.

To understand this, let us consider the case of a joint probability distribution (JPD) of spatially entangled photon pairs measured in the Fourier plane of a lens in a one-dimensional system parameterized by the transverse position $q$. As shown in Figure~\ref{fig:SetupFF}b, the resulting JPD appears as a 2D image.  
For frequency-degenerate photons, momentum conservation in the SPDC process ($p_s + p_i \approx 0$) leads to $q_s + q_i \approx 0$ in the detection plane, resulting in a JPD that is almost perfectly anti-diagonal (blue line in Fig.~\ref{fig:SetupFF}b). In the case of non-degenerate photons, however, the momentum conservation condition becomes $q_s/\lambda_s + q_i/\lambda_i \approx 0$, resulting in a JPD where the information lies along a line tilted by an angle $\alpha = \arctan(\lambda_s/\lambda_i)$ relative to the anti-diagonal (red line in Fig.~\ref{fig:SetupFF}b).
As a result, when projecting the JPD along the sum-coordinate axis - which, in the case of a two-dimensional JPD, corresponds to summing the pixel values along all anti-diagonals - the presence of a tilt introduces an artificial broadening of the correlation peak, leading to an overestimation of the true momentum correlation width $\Delta_p$ (Fig.~\ref{fig:SetupFF}c).
In this work, we adapt the method to project the JPD in the momentum basis based on the photon frequency ratio, thereby compensating for this spectral asymmetry and enabling an accurate estimation of $\Delta_p$.

\section{Experimental methods}\label{sec:Exp} 
Our experimental setup is displayed in Figure~\ref{fig:SetupFF}d. A periodically poled potassium-titanyl-phosphate (ppKTP) crystal is pumped by a narrowband continuous-wave $(405\pm0.05)$\;nm laser. Two crystals with lengths of $5$\;mm and $10$\;mm are alternatively used for the experiments. They are both cut for type-0 SPDC and produce non-degenerate signal and idler photons with central wavelengths of $  \lambda_{i}=730$ nm and $\lambda_{s}=910\;$nm, respectively. After the crystal, a dichroic mirror (DM$_1$) reflects the SPDC photons away from the transmitted pump beam. A first lens $\textrm{L}_\textrm{1}$ ($f=100$\;mm) is positioned at a focal distance from the output surface of the crystal. 
A second dichroic mirror (DM$_2$) separates the signal and idler photon paths. Bandpass filters, $\textrm{BPF}_{\textrm{i}} \sim (730 \pm 1.5)$ nm and $\textrm{BPF}_{\textrm{s}} \sim (910 \pm 10)$ nm, are placed in the each paths. The two beams are then spatially recombined side by side using a half-wave plate (HWP) and a polarizing beam splitter (PBS), and directed onto two distinct regions of an EMCCD camera. 
For position correlation measurements, a second lens, $\textrm{L}_2$ ($f = 300$ mm), is placed after the PBS at a distance equal to its focal length from the camera. This configuration images the output surface of the crystal onto the camera. For momentum correlation measurements, a third lens, $\textrm{L}_3$ ($f = 100$ mm), is inserted between $\textrm{L}_2$ and the camera, also positioned at a distance equal to its focal length from the camera. This arrangement forms the Fourier image of the crystal output surface onto the camera. A lens $\textrm{L}_p$ is positioned before the ppKTP to control the pump waist.  

\begin{figure}[h]
\centering
\includegraphics[width=1\linewidth]{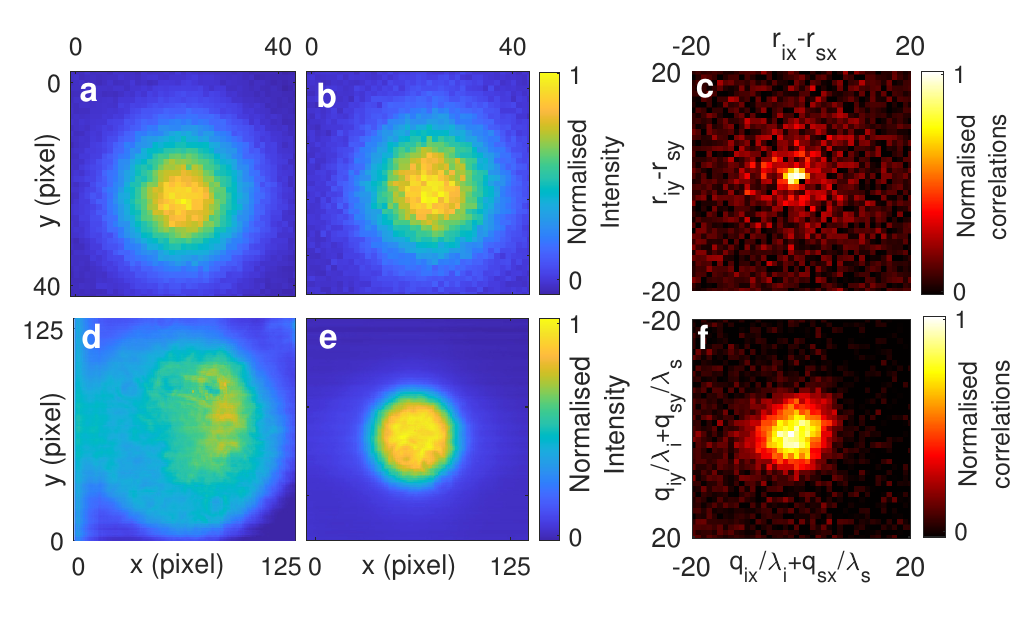}
\caption{\label{Figure2} \textbf{a and b,} Intensity images measured in the imaging configuration for the signal and idler photons, respectively. \textbf{c,} Projection of the spatial joint probability distribution (JPD) along the minus-coordinate axis $\vec{q_s}-\vec{q_i}$. \textbf{d and e,}.  Intensity images measured in the Fourier imaging configuration for the signal and idler photons, respectively. \textbf{f,} Projection of the spatial JPD along the corrected sum-coordinate axis $\vec{q_s}/\lambda_s+\vec{q_i}/\lambda_i$. This data was taken with a crystal with a thickness of $5\;\textrm{mm}$ and a pump waist of $60\;\mu$m.}
\end{figure}

\begin{figure}[h]
\centering
\includegraphics[scale=0.45]{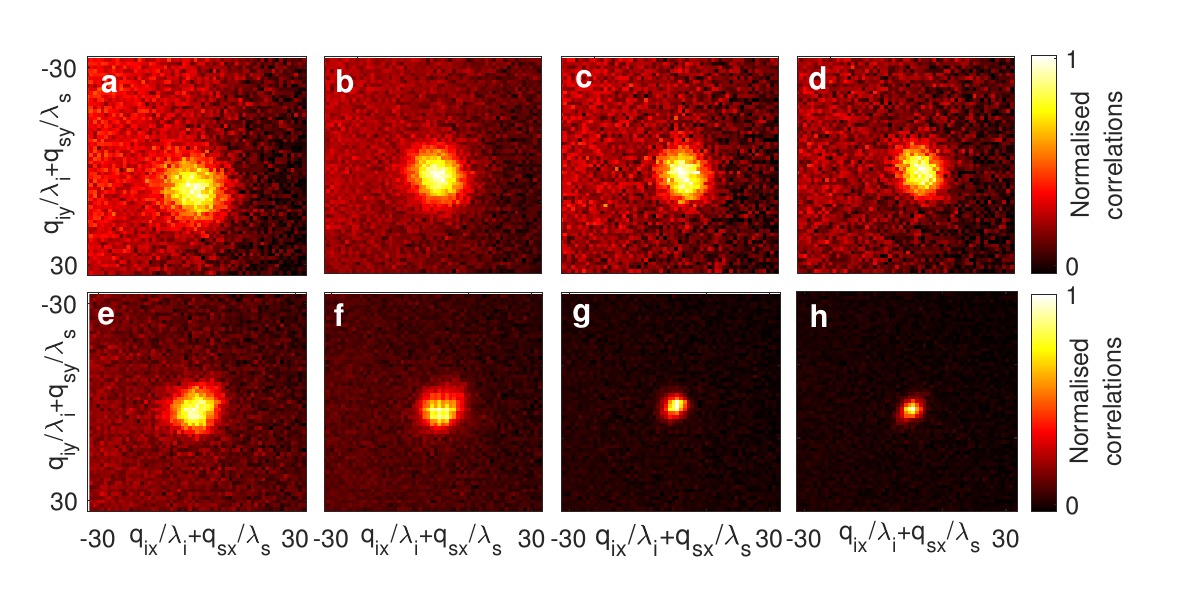}
\caption{\label{fig:Comparison} \textbf{a-d,}  Projections of the spatial joint probability distribution (JPD) along the sum-coordinate axis $\vec{q_s}+\vec{q_i}$ for different pump waists: (a) $60\;\mu$m, (b) $80\;\mu$m, (c) $140\;\mu$m, (d) $160\;\mu$m. \textbf{e-h,} Projections of the spatial JPD along the corrected sum-coordinate axis $\vec{q_s}/\lambda_s+\vec{q_i}/\lambda_i$ for the same pump waists i.e. (e) $60\;\mu$m, (f) $80\;\mu$m, (g) $140\;\mu$m, (h) $160\;\mu$m. These measurements were taken using a crystal with a thickness of $5\;$mm.} 
\end{figure}

\section{Estimation of correlation strength}\label{ImgMeas}

Figures~\ref{Figure2}a and b show the signal and idler intensity images obtained in the imaging configuration (lenses $\textrm{L}_1$ and $\textrm{L}_2$) using nonlinear crystals of $5$ mm thickness and a pump beam waist of $60$$\mu$m. To estimate the position correlation width $\Delta_r$, the spatial JPD, denoted $\Gamma(\vec{r}_s, \vec{r}_i)$, is measured using the EMCCD camera following the method described in Ref.\cite{Defienne}. The projection of the JPD along the minus-coordinate axis ($\vec{r}_s - \vec{r}_i$) reveals a central peak whose width corresponds to $\Delta_r$ (Fig.\ref{Figure2}c). Using a Gaussian fit, we obtain $\Delta_r^{(5\;\textrm{mm})} = 7.5\pm 0.6$~$ \mu$m. 

\begin{figure}
    \centering
\includegraphics[scale=0.43]{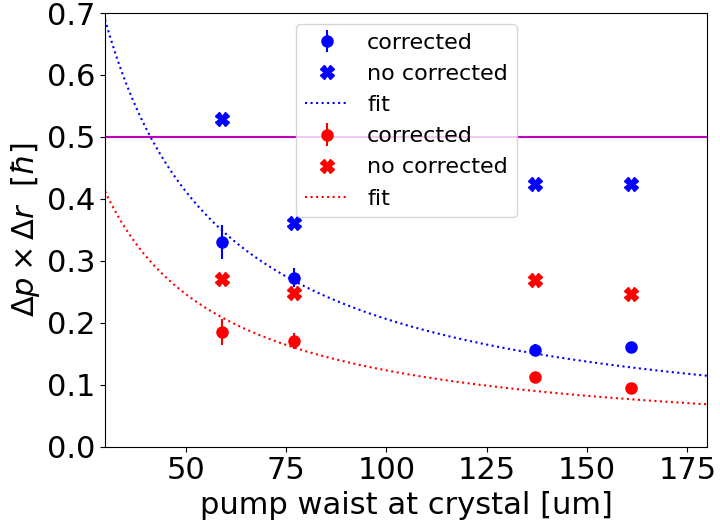}   \caption{EPR products $\Delta_p \Delta_r$ are shown for two nonlinear crystals of different lengths - 5 mm (red) and 10 mm (blue) - as a function of the pump beam waist. Cross markers indicate values computed without correcting the sum-coordinate projections to estimate $\Delta_p$, whereas circular markers indicate values obtained with the correction applied.
The data are fitted using the function $\Delta_p \Delta_r = {a_L}/{w_p}$, where $w_p$ is the pump beam waist and $a_L$ is a fitting coefficient. We find $a_{5\,\mathrm{mm}} = (20 \pm 1)\;\mu\mathrm{m}$ and $a_{10\,\mathrm{mm}} = (12 \pm 1)\;\mu\mathrm{m}$, yielding a ratio $a_{5\,\mathrm{mm}} / a_{10\,\mathrm{mm}} = 1.7 \pm 0.2$, which is close to the theoretically expected value of $\sqrt{2}$~\cite{schneeloch_introduction_2016}. The purple horizontal line correspond to $\Delta_p \Delta_r = 0.5 \hbar$.}
    \label{fig:EPRexp}
\end{figure}

For frequency-degenerate photon pairs, the measurement of $\Delta_p$ follows a similar procedure: the JPD is measured in the Fourier imaging configuration (using lenses $\textrm{L}_1$, $\textrm{L}_2$, and $\textrm{L}_3$) and projected along the sum-coordinate axis ($\vec{q}_s + \vec{q}_i$). For non-degenerate photons, however, this projection overestimates the momentum correlation width, as explained in Figure~\ref{fig:SetupFF}c. 
To account for the frequency difference, the projection is calculated using the modified formula:
\begin{equation}
\label{sumOffdiag}
    \Gamma^+(\vec{q}_{+}^{\alpha}) = \iint \Gamma \left(\lambda_s \vec{q}_+^{\alpha} - \frac{\lambda_s}{\lambda_i} \vec{q}_i,\vec{q}_i \right) d\vec{q}_i,
\end{equation}
where $\vec{q}_{+}^{\alpha}=\vec{q}_s/\lambda_s +  \vec{q}_i/\lambda_i$ is the corrected sum-coordinate variable and $\alpha = \arctan(\lambda_s / \lambda_i) \approx 0.9$.
%where $\vec{q}_{\textcolor{red}{\pm,\alpha}}=\vec{q}_s/\lambda_s \textcolor{red} {\pm} \vec{q}_i/\lambda_i$ is the corrected sum-coordinate variable and $\alpha = \arctan(\lambda_s / \lambda_i) \approx \textcolor{red}{38.7^\circ}$. 
As in the 2D case detailed in Figure~\ref{fig:EPRexp}, \eqref{sumOffdiag} enables projection of the 4D JPD along an axis forming an angle $\alpha$ with respect to the anti-diagonal.
Figures~\ref{Figure2}d and e show the signal and ider intensity images, and Figure~\ref{Figure2}f shows the corrected sum-coordinate projection. Using a Gaussian fit, we extract correlation width values of $\Delta_p^{(60\;\mu \textrm{m})} = (24.6 \pm 0.4) \hbar $~$\textrm{mm}^{-1}$. 

\begin{table*}[h]
\caption{Experimental values of $\Delta_p$, $\Delta_r$, and the corresponding EPR product $\Delta_p \Delta_r$ are reported for two crystal lengths ($5$ mm and $10$mm) under varying pump beam waists. The associated uncertainties are evaluated using the same method described in Ref.~\cite{Defienne}. All listed values correspond to the positions and correlation widths in the crystal plane, obtained after accounting for the magnification and the lenses present in the imaging system.}
    \label{tab:NF_R}
    \centering
    \begin{tabularx}{1.0\textwidth}{|p{1.7cm}||p{1.7cm}|p{1.7cm}|p{1.7cm}||p{1.7cm}|p{1.7cm}|p{1.7cm}|}
    \cline{1-7}
    Pump waist & \multicolumn{3}{c|}{Crystal thickness: 10$\;$mm} &\multicolumn{3}{c|}{Crystal thickness: 5$\;$mm}  \\
\cline{2-7}
$\pm5\;[\mu$m]  & $\Delta_r\;[\mu\textrm{m}]$ & $\Delta_p\;[\hbar/\textrm{mm}]$ & $\Delta_r\Delta_p \;[\hbar]$ & $\Delta_r\;[\mu \textrm{m}]$ & $\Delta_p \;[\hbar/\textrm{mm}]$ & $\Delta_r\Delta_p \;[\hbar]$  \\ 
\cline{1-7}
  60 & 13.9$\pm 0.6$ & 23.8 $\pm 0.4$ & 0.33 $\pm 0.03$ & 7.5$\pm 0.6$ & 24.6$\pm 0.4$ & 0.18 $\pm 0.02$ \\
  
  80 & 13.9$\pm 0.2$ & 19.6$\pm 0.4$ & 0.27 $\pm 0.02$ & 8.0$\pm 0.4$ & 21.3$\pm 0.2$ & 0.17 $\pm 0.01$ \\
  
  140 & 13.9$\pm 0.2$  & 11.3$\pm 0.4$& 0.16 $\pm 0.01$ & 9.6$\pm 0.3$ & 11.7$\pm 0.2$ & 0.11 $\pm 0.01$ \\
  
  160 & 14.3$\pm 0.4$ & 11.3$\pm 0.4$ & 0.16 $\pm 0.01$ & 9.1$\pm 0.5$ & 10.4$\pm 0.6$ & 0.09 $\pm 0.01$ \\
  \cline{1-7}
    \end{tabularx}
\end{table*}

The importance of this angular correction is highlighted in Figure~\ref{fig:Comparison}, which shows the sum-coordinate projections without correction (a–d) and with correction (e–h) for different pump waists. While the uncorrected projections (Figs.~\ref{fig:Comparison}a–d) do not exhibit significant variations in $\Delta_p$, a clear decrease is observed as the waist increases in the corrected projections (Figs.~\ref{fig:Comparison}e–h), as expected in theory~\cite{schneeloch_introduction_2016}.
In particular, the uncorrected projections obtained for pump waists of $60~\mu\textrm{m}$ and $160~\mu\textrm{m}$ yield significantly larger values of $\Delta_p$: $\Delta_p^{(60~\mu\textrm{m})} \approx 36 \hbar~\textrm{mm}^{-1}$ and $\Delta_p^{(160~\mu\textrm{m})} \approx 27\hbar~\textrm{mm}^{-1}$. 
The values of $\Delta_r$ and $\Delta_p$ measured for four different pump waists and two crystal lengths are reported in Table~\ref{tab:NF_R} (see also supplementary document for intensity and correlation images obtained with the $10$mm-thick crystal).

\section{Entanglement certification}
Finally, to certify the presence of entanglement, the products $\Delta_r \Delta_p$ are calculated from the correction width values measured for different crystal lengths and pump waists. These results are reported in Table~\ref{tab:NF_R}. Additionally, Figure~\ref{fig:EPRexp} shows the products obtained with correction (circular markers) and without correction (crosses markers) of the sum-coordinate projections. The values obtained with correction vary inversely with the pump waist size, consistently violating \eqref{eprreid} in a statistically significant manner. In contrast, the uncorrected values show little variation, and one of them (crystal length of $10$ mm and pump waist of $160 \mu$m) is even too large to conclude the presence of entanglement. This case highlights that applying angular correction to the sum-coordinate projection is crucial for certifying entanglement in the case of non-degenerate photon pairs.

\section{Conclusions}

We have investigated position and momentum correlations of non-degenerate photon pairs produced by SPDC using a camera-based coincidence detection technique. Our results confirm that ignoring the wavelength discrepancy significantly distorts the measured momentum correlations and thus undermines the spatial entanglement certification through an EPR criterion. By introducing a corrected angular projection in the JPD analysis, we achieve accurate estimates of the momentum correlation width. Moreover, we show how the relevant parameters - pump waist and crystal length - affect the observed entanglement, in good agreement with theory. This study highlights the importance of appropriately accounting for wavelength mismatches in multi-color SPDC experiments, ensuring reliable spatial correlation measurements and entanglement certification, and facilitating broader applications of multi-wavelength photon pairs in quantum imaging and communication.

\begin{backmatter}
\bmsection{Funding} 
We thank the Fraunhofer Mobility Program from the Fraunhofer Zentrale, that supported the mobility of Emma Brambila for realizing the experiments involved here.
H.D. acknowledges funding from an ERC Starting Grant (grant no. SQIMIC-101039375).

%Finally, the project funds XXX...\textcolor{red}{MISSING? RENE? HUGO? I don't have any that need to be added.}

\bmsection{Acknowledgment} 
We wish to acknowledge the support of colleagues at the INSP, Patrick Cameron, Chloé Vernier, Raj Pandya, and Baptiste Courme, for the discussions related to this project. We also thank Marta Gilaberte-Basset, Valerio Gili and Josué León Torres at the Fraunhofer IOF for valuable discussions.

\bmsection{Disclosures} The authors declare no conflicts of interest.

\bmsection{Authors contributions}
E.B. designed and performed the experiments. E.B. and R.G. analyzed the data. E.B and R.S. conducted the theoretical analysis. All authors discussed the results and contributed to the manuscript. H.D. and M.G. conceived the original idea and supervised the project.

\bmsection{Data availability} Data underlying the results presented in this paper are not publicly available at this time but may be obtained from the authors upon reasonable request.

\end{backmatter}

\bibliography{sample}

\bibliographyfullrefs{sample}

\end{document}